\documentclass[twocolumn,showpacs,preprintnumbers,amsmath,amssymb]{revtex4}
\usepackage{amssymb}
\usepackage{amsmath}
\input{epsf}
\begin{document}

\title{Response to Comment on `Undamped electrostatic plasma
waves' [Phys. Plasmas 19, 092103 (2012)]}
\author{F. Valentini,$^1$ D. Perrone,$^1$  F. Califano,$^2$ F. Pegoraro,$^2$ 
P. Veltri,$^1$ P. J. Morrison,$^3$  and T. M. O'Neil$^4$}
\affiliation{$^1$Dipartimento di Fisica and CNISM, Universit\`a della Calabria,
87036 Rende (CS), Italy \\ $^2$Dipartimento di Fisica and CNISM, Universit\`a di
Pisa, 56127 Pisa, Italy\\ $^3$Department of Physics and Institute for Fusion Studies, University of Texas at
Austin, Austin, TX 78712-1060\\
$^4$Department of Physics, University of California at San Diego, La Jolla,
California, 92093}


\begin{abstract}
Numerical and experimental evidence is given for the occurrence of the plateau states and concomitant corner modes proposed in \cite{valentini12}.  It is argued that these states provide a better description of reality for small amplitude off-dispersion disturbances than the conventional Bernstein-Greene-Kruskal or cnoidal  states such as those proposed in \cite{comment}.
\end{abstract}

\date{\today}
\maketitle

Since the publication of the original  Bernstein-Greene-Kruskal (BGK)  paper \cite{bgk}, which described ways to construct a large class of nonlinear wave states,  there has been an enormous literature that speculates about which of these  states might occur in experiments, in nature,  and in numerical simulations in various situations \cite{bmt}. In his Comment \cite{comment}, Schamel has written a broad spectrum diatribe touching on many points; we agree with some of these points, in fact, some of them were originally advocated by some of us.  In particular, we are in agreement that off-dispersion excitations,  made possible by particle trapping, are important.  However, we will confine our response to those of his comments  that are relevant to our paper, which is about the construction of an appropriate linear theory that describes the small amplitude limit.  Schamel's claim  is that our lack of trapped particles invalidates our analysis even in this limit. Rather than reiterate the points made in our paper,  we  rebut his claim by providing further numerical and experimental evidence.

Our point is best made by a glance at Fig.~\ref{fig4} below that shows late time simulation results for two runs:  an off-dispersion case (Run B),  with  the wide nearly $x$-independent plateau that we  proposed for the distribution function, and an on-dispersion case (Run A), depicting the  more conventional BGK or cnoidal wave state.   Evidently,  the  more conventional BGK or cnoidal wave state is  not the best description of the late time dynamics of the off-dispersion simulation.
\begin{figure}
\epsfxsize=7cm \centerline{\epsffile{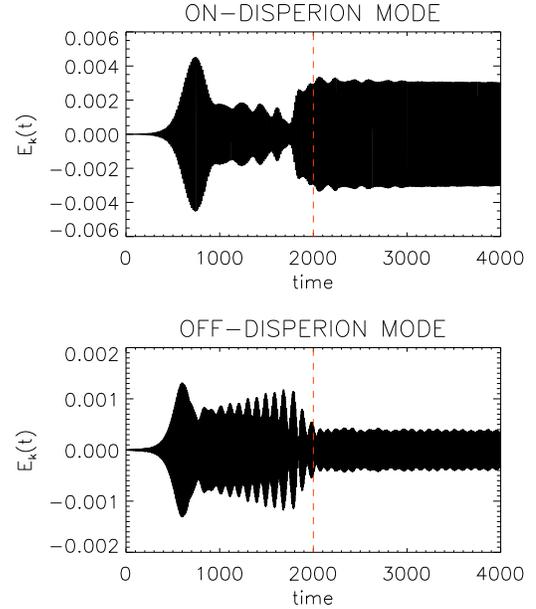}}     
\caption{(Color online) Time evolution of the fundamental spectral component of the electric field $E_k(t)$ of an on-dispersion  mode (Run A at the top) and of an off-dispersion mode (Run B at the bottom). The vertical dashed lines represent $t_{\rm off}$.} \label{fig1}
\end{figure}

In the following we will describe our nonlinear simulations that produced  Fig.~\ref{fig4} and give comparison to our theory of  \cite{valentini12}.  This will be followed by a discussion of some  experimental results that further provide evidence for the validity of our theory and, in particular, the existence of corner modes.

As in Ref. \cite{valentini12}, we make use of a Eulerian code \cite{valentini05,valentini07,valentini07_2} that solves the Vlasov-Poisson equations for one spatial and one velocity dimension:
\begin{equation}
\label{eqVP}
\frac{\partial f}{\partial t}+ v\frac{\partial f}{\partial
x}-E\frac{\partial f}{\partial v}  = 0\,, \qquad  \frac{\partial
E}{\partial x}= 1-\int f dv\,,
\end{equation}
where $f=f(x,v,t)$ is the electron distribution function and $E=E(x,t)$ the electric field. In (\ref{eqVP}), the ions are a neutralizing background of constant density $n_0=1$, time is scaled by the inverse electron plasma frequency $\omega_p^{-1}$, velocities by the electron thermal speed $v_{th}$,  and lengths by the electron Debye length $\lambda_D$.  For   simplicity,  all the physical  quantities will be expressed in these characteristic units. The phase space domain for the simulations is $\mathcal{D}=[0,L]\times[-v_{_{max}},v_{_{max}}]$. Periodic boundary conditions in $x$ are assumed, while the electron velocity distribution is set equal to zero for $|v|>v_{_{max}}=6$. The $x$-direction is discretized with $N_x=256$ grid points, while the $v$-direction with  $N_v=12001$.

In our   previous simulations of Ref.~\cite{valentini12},  the initial equilibrium consisted of a  velocity distribution function  with a small plateau; however,  here we assume a plasma with an initial Maxwellian velocity distribution and homogeneous density.  We then use an external driver electric field that can  dynamically  trap resonant electrons and  create a plateau in the velocity distribution. This is the same approach used in the numerical simulations of Ref.~\cite{afeyan04,valentini06,johnston09} and in the experiments with nonneutral plasmas in Ref.~\cite{anderegg09}.

The explicit form of the external field is
\begin{equation}\label{driver}
 E_{_D}(x,t)=g(t)E_{_{DM}}\sin{(kx-\omega_{_D} t)}\,,
\end{equation}
where $E_{_{DM}}$ is the maximum driver amplitude, $k=2\pi/L$ is the drive wavenumber with $L$  the maximum wavelength that fits in the simulation box, $\omega_{_D}=kv_{\phi_{_D}}$ is the drive frequency with $v_{\phi_{_D}}$ the driver phase velocity, and $g(t)=[1+(t-\tau)^n/\Delta\tau^n]^{-1}$ is a profile that determines the ramping up and ramping down of the drive.  The external electric field is applied directly to the electrons by adding $E_{_D}$ to $E$ in the Vlasov equation. An abrupt turn-on or turn-off of the drive field would excite Langmuir (LAN) waves and complicate the results. Thus, we  choose $n=10$  so  $g(t)$  amounts to a nearly adiabatic turn-on and turn-off. The driver amplitude remains near $E_{_{DM}}$ for a time interval of order $\Delta\tau$ centered at $t=\tau$ and  is zero for $t\geq t_{\rm off}\simeq\tau+\Delta\tau/2$. We will analyze the  plasma response for many wave periods after the driver has been turned off.
\begin{figure}
\epsfxsize=7cm \centerline{\epsffile{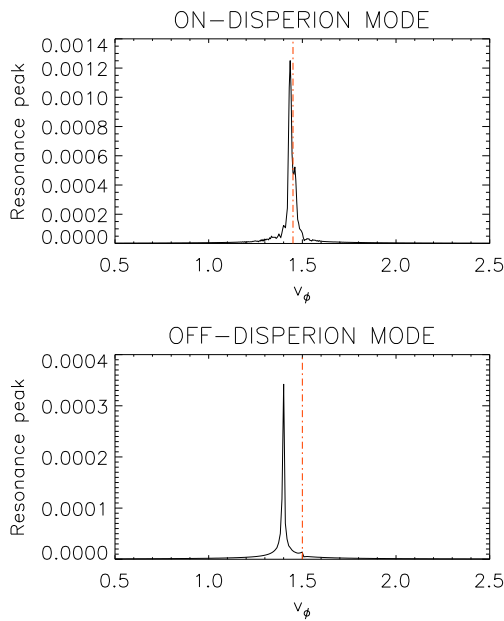}}        
\caption{(Color online) Resonance peak for Run A (top) and Run B (bottom); the red-vertical lines indicate the value of the driver phase velocity $v_{\phi_{_D}}$ for Run A and B,  respectively.} \label{fig2}
\end{figure}

We  simulate both  the excitation of an on-dispersion  mode (Run A),   a mode for which $(k,v_\phi)$ is  on the thumb curve of  Fig.~1 of Ref.~\cite{valentini12}, and  an off-dispersion mode (Run B), that  is off the thumb curve. The excitation of the on-dispersion mode is obtained through an external driver with $k=\pi/10$ and $v_{\phi_{_D}}=1.45$, while for the off-dispersion mode we set $k=0.7$ and $v_{\phi_{_D}}=1.5$. The maximum driver amplitude $E_{_{DM}}=0.01$ has been chosen for each simulation in such a way that $\Delta\tau\simeq 10 \tau_t$, with $\tau_t$ being the trapping period \cite{oneil65}. Finally, the maximum time of the simulation is $t_{max}=4000$, while the driver is zero for $t\geq t_{\rm off}=2000$.
\begin{figure}
\epsfxsize=7cm \centerline{\epsffile{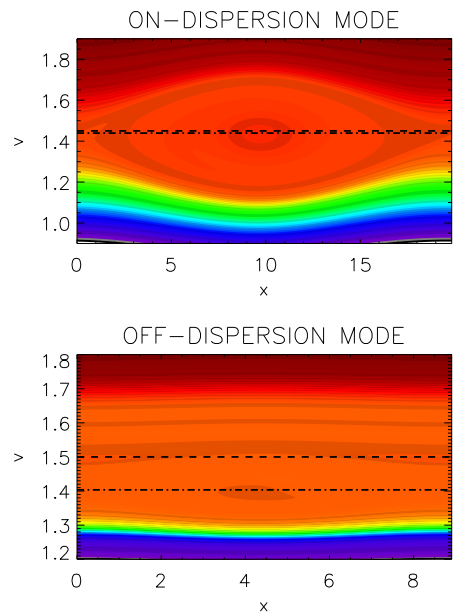}}   
\caption{(Color online) Phase space contour plot of the electron distribution function at $t=t_{max}$ for Run A (top) and Run B (bottom). Black dashed and black dot-dashed lines indicate the driver phase speed $v_{\phi_{_D}}$ and the mode phase speed $v_\phi$, respectively.} \label{fig3}
\end{figure}

Figure \ref{fig1} shows the time evolution of the fundamental spectral component  of the electric field, $E_k(t)$,  for Run A (top) and Run B (bottom).  In both plots one can see that after the driver has been turned off at $t_{\rm off}$ (indicated by the red-dashed lines in the figures),  the electric field oscillates at a nearly constant amplitude.
\begin{figure}
\epsfxsize=8cm \centerline{\epsffile{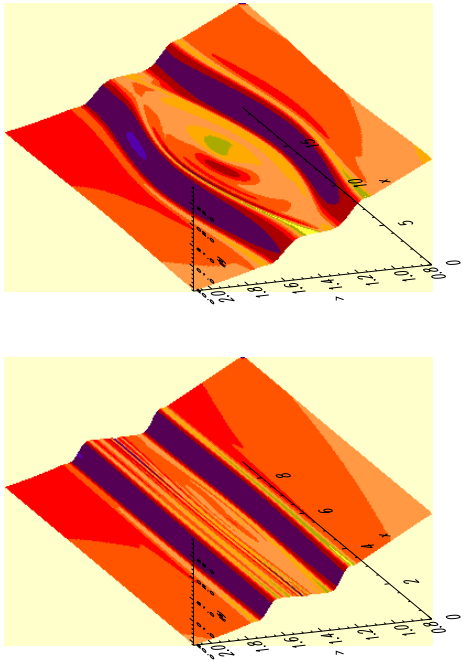}}   
\caption{(Color online) Surface plot of the electron distribution function at $t=t_{max}$ for Run A (top) and Run B (bottom).} \label{fig4}
\end{figure}
Figure \ref{fig2} depicts  the resonance peaks for Run A (top) and Run B (bottom), obtained through  Fourier analysis of the numerical electric signals performed in the time interval $t_{\rm off}\leq t\leq t_{max}$, i.e.,  in the  absence of the external driver. From these two plots it is readily seen that the  on-dispersion mode propagates with phase speed very close to the driver phase velocity $v_{\phi_{_D}}$, whose value is indicated by the vertical red-dashed lines, while the phase speed of the  off-dispersion mode is shifted towards a lower value  with respect to $v_{\phi_{_D}}$. This shift is predicted by our theory of Ref.~\cite{valentini12} and these results provide qualitative evidence for its validity; subsequently, we will show  quantitative agreement.

The main differences between the  on-dispersion and off-dispersion modes can be appreciated by examination of  the phase space contour plots of the electron distribution function shown at $t=t_{max}$ in Fig.~\ref{fig3}.  For  Run A (top) a well defined trapping region  that propagates in the positive $x$-direction is visible. The black-dashed and black dot-dashed lines in the figure represent the phase speed of the driver $v_{\phi_{_D}}$ and  excited mode $v_\phi$, respectively.  For  Run A it is easily seen that $v_{\phi_{_D}}$ and $v_\phi$ are almost identical, meaning that the region of trapped particles generated by the external driver survives even when the driver is off and that this region  streams with a mean velocity close to that of the  driver.  The physical scenario appears quite different for the off-dispersion mode of Run B (bottom).  Here we observe a rather wide nearly $x$-independent plateau (the orange region of the plot) that is substantially wider than the separatrix for the trapped particles (small dark region at velocity close to $v\simeq 1.4$). Moreover, here the values of $v_{\phi_{_D}}$ and $v_\phi$ are well-separated, meaning that the excited mode oscillates with a frequency smaller than that of the external driver.

\begin{figure}
\epsfxsize=7cm \centerline{\epsffile{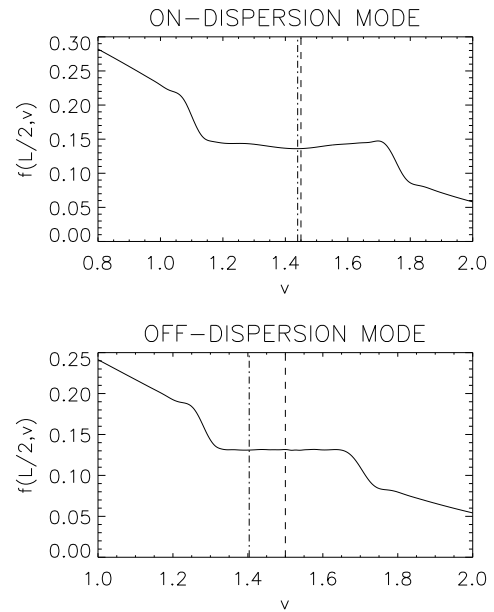}}     
\caption{Velocity dependence of $f(L/2,v)$ at $t=t_{max}$ for Run A (top) and Run B (bottom); black dashed and black dot-dashed lines indicate $v_{\phi_{_D}}$ and $v_\phi$,  respectively} \label{fig5}
\end{figure} 

The differences between Run A and Run B can be further appreciated by looking at the electron distribution function surface  plots of Fig.~\ref{fig4}. For Run A (top) we observe a trapped region modulated in the spatial direction, while for
 Run B (bottom) we see a flat region whose velocity width appears to be independent of $x$.   The on-dispersion  Run A resembles the more conventional  BGK type solution like Schamel's,   while  the nearly $x$-independent off-dispersion plateau of Run B is very different, it being more like a quasilinear plateau.  Since the  on-dispersion case has no frequency shift,  it appears that  the trapping dynamics is dominated by a single wave and a  BGK type solution is to be expected.  For the off-dispersion case, where there is a frequency shift between the driver and the ringing wave, the trapping dynamics may involve multiple waves with  different phase velocities.  The interaction between these waves could be causing a band of chaotic dynamics that re-arranges phase space to provide a more quasilinear type of plateau.

 Quantitative evidence for our theory can be extracted from Fig.~\ref{fig5}, which shows  $f(x=L/2,v)$ as a function of $v$ for Run A (top) and Run B (bottom).  Again, black dashed and black dot-dashed lines indicate $v_{\phi_{_D}}$ and $v_\phi$,  respectively. Also here the phase velocity shift for the off-dispersion mode is evident; by taking into account the uncertainty due the finite time resolution of the simulations, we can estimate the interval in which the value $\Delta v_\phi^{(num)}:=v_\phi-v_{\phi_{_D}}$ of the phase speed shift falls.  This gives 
 $-0.095<\Delta v_\phi^{(num)}<-0.0104$. Using this we can compare the phase velocity shift obtained for Run B to the analytical prediction using the  ``rule of thumb" of Eq.~(23) in Ref.~\cite{valentini12}: the theoretical expectation for the phase velocity shift of the off-dispersion  mode of Run B is $\Delta v_\phi^{(th)}\simeq -0.0946$ (with a value $\Delta v_\phi^{(th)}\simeq -0.0933$ obtained by increasing the resolution by two order of magnitude in velocity), in very good agreement with the value obtained from the simulation.  Thus our theory not only predicts the qualitative direction of the phase velocity shift, it gives a very good quantitative value. 
 
To conclude, we show  that there already exists published  qualitative experimental evidence \cite{anderegg09} for the validity of the theory we proposed in \cite{valentini12}.    Figure~6  of Ref.~\cite{anderegg09} depicts the  plasma response to a drive at a spread of frequencies.  In Fig.~6(b) the  larger peak on the right corresponds to the Trivelpiece-Gould mode, which in this experiment corresponds to the LAN mode of the thumb curve, while the small peak on the right corresponds to the Electron-Acoustic wave. Thus, frequencies between these peaks and above the LAN peak correspond to off-dispersion modes.  In addition,  at the bottom of Fig.~6(b) is given an indication of the frequency shift between  the plasma response,  corresponding  to a ringing mode,  and  that of the drive.  Since $k$ is fixed, this frequency shift is equivalent to a shift in the phase velocity.  Observe that the frequency shift is positive within the thumb curve (between the peaks) and is negative above the LAN peak The directions of these shifts can be inferred from  our theory, cf.\   the rule of thumb, Eq.~(23) of Ref.~\cite{valentini12}.   In this equation  $k^2 = M$   gives a frequency on the thumb curve, while a frequency of a corner mode, an off-dispersion excitation,  is obtained by seeking a root with the addition of the plateau contribution (the remaining term of Eq.~(23)).  Because  the rule of thumb gives the local shape of the plateau contribution,  it is not difficult to infer the direction of the frequency shift relative to the drive.  A straight bit of reasoning using the rule of thumb  implies frequencies within the thumb curve and those above the LAN mode should shift in precisely  the directions  seen in the experiments.
 
More details about the simulations discussed here  and further experimental verification of our theory will be the subjects of future works  that are presently under preparation.  

{\it Postscript:  In his response to the first version of the present Response, Schamel  modified his Comment  \cite{comment} in an attempt  to use our numerical and experimental  results to substantiate his case.   We are not convinced by his arguments and stand by our  original conclusion of \cite{valentini12}; viz.,  corner modes,  under the circumstances we described, provide a better description of computational and experimental results than cnoidal/BGK modes.}

\section*{Acknowledgments}

We thank Prof.\ Schamel for providing this opportunity to further substantiate the validity of our results.
The numerical simulations were  performed on the FERMI supercomputer at CINECA (Bologna, Italy), within the European project PRACE Pra04-771.  P.J.M.\ was supported by  Department of Energy grant  DEFG05-80ET-53088. T.M.O.\  was supported by National Science Foundation grant PHY-0903877 and Department of Energy grant DE-SC0002451.


\end{document}